# Characterization of photoinduced normal state through charge density wave in superconducting YBa$_2$Cu$_3$O$_{6.67}$

H. Jang,[1,2,†] S. Song,[3,†] T. Kihara,[4] Y. Liu,[5] S.-J. Lee,[5] S.-Y. Park,[1] M. Kim,[1] H.-D. Kim,[1] G. Coslovich,[3] S. Nakata,[6] Y. Kubota,[7,8] I. Inoue,[7] K. Tamasaku,[7] M. Yabashi,[7,8] H. Lee,[9] C. Song,[2,9] H. Nojiri,[4] B. Keimer,[6] C.-C. Kao,[10] and J.-S. Lee[5,*]

[1]PAL-XFEL, Pohang Accelerator Laboratory, Pohang, Gyeongbuk 37673, Republic of Korea

[2]Photon Science Center, Pohang University of Science and Technology, Pohang, Gyeongbuk 37673, Republic of Korea

[3]Linac Coherent Light Source, SLAC National Accelerator Laboratory, Menlo Park, California 94025, USA

[4]Institute for Materials Research, Tohoku University, Katahira 2-1-1, Sendai, 980-8577, Japan

[5]Stanford Synchrotron Radiation Lightsource, SLAC National Accelerator Laboratory, Menlo Park, California 94025, USA

[6]Max Planck Institute for Solid State Research, Heisenbergstr. 1, 70569 Stuttgart, Germany

[7]RIKEN SPring-8 Center, Sayo, Hyogo 679-5148, Japan

[8]Japan Synchrotron Radiation Research Institute, Sayo, Hyogo, 679-5198, Japan

[9]Departments of Physics, Pohang University of Science and Technology, Pohang, Gyeongbuk 37673, Republic of Korea

[10]SLAC National Accelerator Laboratory, Menlo Park, California 94025, USA

[†]These authors contributed equally to this work.

[*]Correspondence to: jslee@slac.stanford.edu (Jun-Sik Lee)

**The normal state of high-Tc cuprates has been considered one of the essential topics in high-temperature superconductivity research. However, compared to the high magnetic-fields study of it, understanding a photoinduced normal state remains elusive. Here, we explore a photoinduced normal state of YBa$_2$Cu$_3$O$_{6.67}$ (YBCO) through a charge density wave (CDW) with time-resolved resonant soft x-ray scattering, as well as a high-magnetic field x-ray scattering. In the non-equilibrium state in which people predict a quenched superconducting state based on the previous optical spectroscopies, we experimentally observed a similar analogy to the competition between superconductivity and CDW shown in the equilibrium state. We further observe that the broken pairing states in the superconducting CuO$_2$ plane via the optical pump lead to nucleation of three-dimensional CDW precursor correlation, revealing that the photoinduced CDW is similar to phenomena shown under magnetic fields. Ultimately, these findings provide a critical clue that the characteristics of the photoinduced normal state show a solid resemblance to those under magnetic fields in equilibrium conditions.**

**Introduction**

High-$T_c$ cuprate is one of the most studied systems in strongly correlated materials since the first discovery of high-temperature superconductivity (HTSC) was reported in 1986 (1). Despite this high level of focus, a fundamental mechanism of HTSC is unclear because of many conundrums (2–4). Even with the tremendous research efforts in this field, realizing room-temperature superconductivity under the ambient pressure remains a huge challenge. Meanwhile, it has been considered that understanding the normal states of high-$T_c$ cuprates and their corresponding features, e.g., pseudogap (5,6) and charge density wave (CDW) (7–20), could shed light on a direction to approach room temperature superconductivity. In this regard, an in-depth study on the ground state of the normal state is of profound significance to HTSC research.

An experimental study on the ground state problem is generally accompanied by the cooling of samples because they should be at the lowest possible energy state. In the HTSC case, due to the emergence of superconductivity (SC) below $T_c$, however, it is difficult to resolve the normal state at low temperature. Typically, this difficulty is overcome by quenching the superconducting state at low temperature by an external magnetic field (21–30). For example, x-ray scattering studies on YBa$_2$Cu$_3$O$_{6+\delta}$ cuprates under high-magnetic fields revealed that the three-dimensional (3D) CDW correlation is the ideal ordering state in the normal state (24–30). Meanwhile, quenching the SC through the optical pump is considered another approach and has been demonstrated in many cuprates (31–39). However, a study on corresponding a normal state triggered by the laser-driven quench of the superconducting state is still in the early stage, although a photoinduced CDW state was recently reported on YBa$_2$Cu$_3$O$_{6+\delta}$ (40). In addition, findings through the pump-probe approach and corresponding implications represent fundamental physics of the non-equilibrium state, while the magnetic field study secures the equilibrium state. In this manner, despite that these two approaches' successful implementations in many high-$T_c$ cuprates, it is still premature to reach a common ground on the normal state studied through those two approaches. Hence, it is timely to establish a scientific connection between the equilibrium state and the non-equilibrium state in HTSC.

Here we explore a photoinduced normal state of ortho-VIII YBa$_2$Cu$_3$O$_{6.67}$ (YBCO) by using a time-resolved resonant soft x-ray scattering (tr-RSXS) study and also carry out a pulsed high magnet field x-ray scattering experiment (see Methods). For this purpose, we focus on investigating CDW phenomena of YBCO while the superconducting state is quenched by the optical pump. This is because we are motivated by the fact that the CDW correlation is a normal state phenomenon (3,4) and has ubiquitously been established in high-$T_c$ cuprates (7–20). In particular, there exists previous extensive previous studies, even with magnetic fields, on the CDW phenomena in YBa$_2$Cu$_3$O$_{6+\delta}$ cuprates (12–14,24–30,40). Thus, it allows us to compare this study to the CDW phenomena measured with the magnetic field, leading to an in-depth insight into the connection between the equilibrium and non-equilibrium states in the high-$T_c$ cuprates. Moreover, there is no systematic experimental study regarding intertwined phenomena between the photoinduced CDW and the quenched SC state. In this study, we directly observe that both intensity and correlation length of the CDW order in YBCO are enhanced while the optical pump suppresses the superconducting state, indicating that the competition between CDW and SC shows a strong resemblance to that under those magnetic fields (13,24–30). It is also worthwhile to note that we basically assume a quench of the SC state through the previous optical demonstrations (31–39). We further observe an emergence of a three-dimensional CDW precursor correlation (27) under the photoinduced normal state of YBCO. From these findings, we bring up a point that CDW phenomena in the equilibrium state are characteristically identical to those in the non-equilibrium state. In addition, we suggest that the broken Cooper-pairing physics could be discussed in the context of the nucleation of the vortex state through the external magnetic field. Inversely, it allows us to extend our scientific consideration even under the equilibrium state into the fact that the non-equilibrium approach could generate a transient superconducting state nearly up to room temperature (41–44).

**Results**

***CDW of ortho-VIII $YBa_2Cu_3O_{6.67}$ under the equilibrium state.*** One of the profound discoveries in high-Tc cuprates is the universal existence of a static CDW (or charge order) in $CuO_2$ planes (7–20). In the underdoped $YBa_2Cu_3O_{6+\delta}$ cuprates, a CDW typically appears in the pseudogap phase (12–14,24–30). In other words, the onset temperature of CDW ($T_{cdw}$) exists in the normal state. Many researchers believe its ground state is critical to understand the superconducting behavior in $YBa_2Cu_3O_{6+\delta}$ (3,28).

Figure 1A shows a temperature dependence of static CDW of the YBCO crystal, which was measured by resonant soft x-ray scattering (RSXS) measurement (see Methods). We observed CDW peaks at $Q = (0, q_{cdw}, \sim1.45)$ where $q_{cdw}$ is around -0.32 reciprocal lattice unit (r.l.u.) in the $CuO_2$ plane (fig. S1). The CDW appears below ~150 K, which is consistent with previous studies (24–27). As expected, the CDW, which is developed from the normal state, is suppressed while the superconducting state emerges below $T_c = 65.5$ K. This suppression demonstrates the existence of competition between CDW and SC in YBCO, as shown in other $YBa_2Cu_3O_{6+\delta}$ cuprates (12–14,24–30). At the same time, it reveals that a study of CDW's ground state at low temperature (i.e., the normal state) is challenging because of the coexistence of SC and its competition with the CDW. Hence, to investigate a low-temperature state of CDW, the SC needs to be quenched. As a magnetic field causes a vortex state (i.e., quenched SC state), an optical pump can also break Cooper pairs in superconducting $CuO_2$ planes (31–40) when the pumping conditions are appropriately tuned (see Fig. 1B).

***CDW of YBCO under the non-equilibrium state.*** Figure 2A shows an experimental configuration of tr-RSXS, aiming to explore CDW phenomena of YBCO at low temperatures.

We used a 1.55 eV (800 nm) optical laser with a fluence of 15 µJ/cm$^2$ (see Methods), aimed to minimize thermal heating (Supplementary). Figure 2B shows the temporal behavior of the CDW order at $T$ = 25 K, which was measured at $Q$ = (0, -0.32, 1.45). Before the pump – i.e., time delay ($\Delta t$) < 0, it shows a persistent CDW signal suppressed by the competition with SC. After the pump, we observed an enhancement of the CDW for 0 < $\Delta t$ < ~7 ps, which is consistent with the previous work (40). For the 0 < $\Delta t$ < 0.5 ps range, the CDW suppression is associated by a photoexcited charge-carrier while SC is suppressed (see Fig. S6) (35). Also, the maximum enhancement of the CDW occurs at $\Delta t$ ~ +0.9 ps. Figure 2C shows the CDW peak profiles at $\Delta t$ < 0 and $\Delta t$ = +0.9 ps. The photoinduced CDW peak at $\Delta t$ = +0.9 ps becomes stronger and sharper than that at the negative time delay ($\Delta t$ = -1.1 ps). This implies that the CDW volume fraction in CuO$_2$ planes at the pumped moment is recovered from that suppressed by the SC.

In order to scrutinize the photoinduced phenomena, we explored the temperature dependence of CDW scanned at $Q$ = (0, $k$, ~1.45). Figure 3A shows the CDW peak profiles measured at both $\Delta t$ = -1.1 ps and +0.9 ps with different sample temperatures. Note that full delay scans are in fig. S2. Below $T_c$, the photoinduced behavior becomes more pronounced. This indicates the CDW enhancement behavior is associated with the quenched SC state. For more quantitative analysis, we fitted both peak profiles and plotted a summary [CDW area (i.e., integrated intensity), peak width, and area difference] in Fig. 3B. The enhanced CDW area clearly shows a contrast below $T_c$. The correlation length, $\xi$ = 2/FWHM (where FWHM is the full width at half maximum of the Lorentzian profile) of the photoinduced CDW peak at $T$ = 15 K is estimated to be ~ 88 Å, which is about a 60 % increase compared to the case at $\Delta t$ = -1.1 ps. Under the scattering perspective, this deduces that the domain size of CDW becomes larger (i.e., its volume fraction) while SC's volume fraction could be suppressed by the pump. These findings show a strong resemblance to CDW phenomena under external magnetic fields (24–30). Moreover, the optically pumped CDW area is maximized at $T$ ~ 55 K. This indicates that the Tc under the optically pumped condition is shifted lower because the superconducting state needs more energy to form under this pumped condition, which is consistent with typical behavior of a superconductor under a magnetic field (22, 45–46). However, the photoinduced CDW peak measured at 65 K (~ $T_c$) is slightly weaker (~17.7 %) than the CDW peak measured at $\Delta t$ = -1.1 ps. It means that optical pumping causes melting of not only SC in the CuO$_2$ planes but also some portion of the CDW regardless of the existence of SC, like the case of pure CDW material (35). Nevertheless, it is possible to observe the photoinduced CDW behavior in YBCO because the temporal dynamics of CDW are slightly faster (about 100 fs) than that of SC (see fig. S4). Hence, the photoinduced CDW features observed at Δt = +0.9 ps could be attributed to a combined response from both the melted portion by optical pumping and recovered portion by photoinduced suppression of SC.

***Three-dimensional CDW correlation in the photoinduced normal state of YBCO.*** Considering the temperature dependence on the photoinduced CDW behavior in YBCO, it is reasonable to find a crossover in CDW phenomena between the optical pump approach and the magnetic field ($H$) approach as a next step. This triggered us to explore a signature of the long-ranged three-dimensional (3D) CDW under the normal state that occurs when an $H$-field stronger than 50 % of $H_{c2}$ is applied (27–29). In this sense, we performed $L$-dependence measurement of the CDW under the transient state, aiming to explore whether the optical pump also induces a 3D CDW behavior in YBCO.

Figure 4A shows CDW peak profiles at both Δt = -1.1 ps and +0.9 ps with different detector (2θ) angles (i.e., $L$-dependence) at $T$ = 25 K. Note that details of this detector angle dependent measurement are in the Supplementary materials. At Δt = -1.1 ps, the CDW peak gets weaker when the $l$-value moves toward 1 r.l.u. (see summarized CDW peak heights in Fig. 4B), which reproduces the typical quasi-2D CDW behavior in $YBa_2Cu_3O_{6+\delta}$ cuprates that the CDW intensity is maximized around a half-integer ($l \sim 1.5$ r.l.u.) (12–14, 24–30). For Δt = +0.9 ps, we observed a different photoinduced behavior with varying 2θ (also see fig. S3). The photoinduced CDW behavior gets pronounced as the l-value decreases from ~1.45 to ~1.08 r.l.u. Nevertheless, no clear feature of a long-ranged 3D order was observed that would have been clearly visible in this YBCO under the high magnetic field (see Fig. 4C). However, as shown in Fig. 4B summary, we found an interesting tendency that the photoinduced CDW order is maximized at $l \sim 1.2$ r.l.u., which is close to an integer $l$-value. This $L$-dependence of the photoinduced CDW indicates a 3D CDW precursor correlation observed in high magnetic field measurements (27). It would also be worthwhile to note that the photoinduced state is still a mixture between the transient state and the reduced SC portion because of not reaching a full 3D CDW order.

According to the magnetic field experiments on YBa$_2$Cu$_3$O$_{6+\delta}$ cuprates (12–14,24–30), the development of the 3D CDW precursor suggests that the SC quenching effect via our optical pump condition is equivalent to that of the external magnetic field effect of between $H$ = 10 and 14 T. Note that this field estimation is the simplest scaling analysis based on the assumption that the primary mechanism of CDW formation is electronic (see fig. S8). Although the lattice is always strongly coupled to an electron and somewhat affects the dynamics when an optical pump induces a thermal effect, the photoinduced CDW order's $L$-dependence (Fig. 4B and fig. S3), which cannot be explained by the lattice change, supports that the thermal heating is not dominant in this case. In this regard, the $T_c$ shift shown in Fig. 3 is also interpreted to correspond to the equilibrium state under the estimated magnetic field (24,25). Furthermore, considering some portion of the CDW in the CuO$_2$ planes melted by the optical pump, the practical CDW portion in the optical pump is smaller than in the magnetic field approach. Thus, it allows us to reconcile why the observed photoinduced effect corresponds to the relatively weak $H$-field.

**Discussion**

Using the tr-RSXS study on the CDW behavior in YBCO, we explore the intertwined phenomena between the CDW and SC in the normal state induced by the optical pump. We reveal that the CDW characteristic behaviors in the non-equilibrium state show close similarity with those under the magnetic field. Clearly, these findings could provide a clue that the transient state characteristics show a solid analogy to the normal state under equilibrium conditions demonstrated by the magnetic field.

Beyond this, we further discuss the implications of our finding. The fact that CDW and/or spin density wave (SDW) orders in high-$T_c$ cuprates are enhanced by the external magnetic field (24,26-28,47–49) indicates that the vortex state is related to the quenched SC by $H$-field. Such a circumstance is depicted as cartoons in Figs. 5A and 5B (top panel). As the external magnetic field is increased, the number of vortices corresponding increases, leading to an increase in the fraction of the quenched SC volume (i.e., increase in the normal state fraction) in the CuO$_2$ plane (49). In the same manner, we depict that the Cooper-paired SC states (Fig. 5A) are transformed into the broken/weakened Cooper pairs through a proper optical pump (Fig. 5B, bottom panel) (33–40). For a short while, it leads to a transient state that becomes dominant over the reduced

SC portion in the CuO$_2$ plane. After all, the two approaches through the optical pump and the external magnetic field drive the quenched SC state, and their corresponding CDW responses are characteristically identical to the normal state. In this sense, we infer that an essential physics in the non-equilibrium state would be in a crossover with the equilibrium physics. This inference suggests that expanding the vortex liquid state by increasing the magnetic field could be alternatively interpreted by physics of the broken Cooper-pairing in the CuO$_2$ plane (see Fig. 5C). Ultimately, it implies that demonstrating the quenched SC state with either approach could be delivering the equivalent physics in HTSC. Inversely, with the equilibrium state, it is considerable that optical pump approaches are possible to drive room temperature transient superconductivity (41–44). Moreover, understanding such a discussion presents a rich opportunity for detailed experimental investigations (e.g., CDW study through synchronizing three pulses between x-ray, an optical pump, and magnetic field) and theoretical interpretations in the future.

## Materials and Methods

***Sample preparation.*** Single crystals of YBa$_2$Cu$_3$O$_{6.67}$ ($T_c$ = 65.5 K) were grown using a flux method (50), and the oxygen content was adjusted by annealing under well-defined oxygen partial pressure. The sample was mechanically detwinned by heating under uniaxial stress (50 ~ 60 MPa was applied at 400 °C). The dimensions of the single crystal were 2.10 mm × 1.05 mm × 0.38 mm along the crystallographic *a*, *b*, and *c*-axis, respectively (lattice parameters: $a$ ~ 3.82 Å, $b$ ~ 3.88 Å, and $c$ ~ 11.725 Å). The single crystal was thoroughly characterized by measuring the static CDW signal at the Stanford Synchrotron Radiation Lightsource (SSRL) beamline 13-3, using Cu $L_3$-edge resonant soft x-ray scattering (RSXS) measurements (fig. S1).

***tr-RSXS measurement.*** All tr-RSXS experiments were carried out at the SSS-RSXS endstation of PAL-XFEL (51). The sample was mounted on a 6-axis open-circle cryostat manipulator and a base temperature of ~15 K using liquid helium. The sample surface was perpendicular to the crystalline c-axis, and the horizontal scattering plan was parallel to the *bc*-plane. X-ray pulses with an ~80 fs pulse duration and a 60 Hz repetition rate were used for the soft x-ray probe. The x-ray was linearly and horizontally polarized (π-polarization) and the photon energy was tuned around the Cu $L_3$-edge (932 eV). A 1.55 eV (800 nm) optical laser pump with a ~50 fs pulse duration and *p*-polarization was provided by a Ti:sapphire laser (see Fig. 2A and fig. S7). It is commonly appreciated that the laser pulses at 800 nm are able to strongly perturb superconductivity in cuprates (33–35, 39, 40, 52–55). The optical laser was nearly parallel to the incident x-ray beam as the angle difference is less than 1°. A pump fluence of 15 µJ/cm$^2$ was mainly used (see fig. S5). The time delay between pump and probe pulses was controlled by a mechanical delay stage. The spot size of the x-ray at the sample position was 100 (H) × 200 (V) µm$^2$ in FWHM and the spot diameter of optical laser was ~500 µm in FWHM. The x-ray scattering signal was collected by an avalanche photodiode (APD), which was enclosed in an aluminum cage, and the front was covered by a 300-nm-thick aluminum filter to ensure light tightness. All signals were recorded by high-speed digitizers and analyzed on a shot-to-shot basis.

***X-ray scattering measurement with a high magnetic field.*** The high field pulsed-magnet experiments were performed at the BL2-EH3 hutch of the SACLA (56). The YBCO sample was attached on a sapphire rod. The rod was mounted on a closed-cycle cryocooler with a capability of a sample base temperature ~10 K. X-ray pulses with a 7 fs pulse duration and a 30 Hz repetition rate were used. The x-ray was linearly and horizontally polarized (π-polarization) and the photon energy was tuned to 8.8 keV, just below the Cu *K*-edge to reduce the fluorescence background (26). X-ray pulse arrived at the top of ~ 1 ms duration pulsed magnetic field with ~30 Tesla. To obtain zero-field background, 10 shots of x-ray only scattering signal were obtained before and after the shot with a magnetic pulse. As shown in Fig. 4c, therefore, the noise levels between *H* = 0 T (80 shots accumulated) and 30 T (4 shots) appear differently (see fig. S9). A half-mega pixel Multi-Port Charge-Coupled Device (MPCCD) was used to capture the 2D scattering profile.

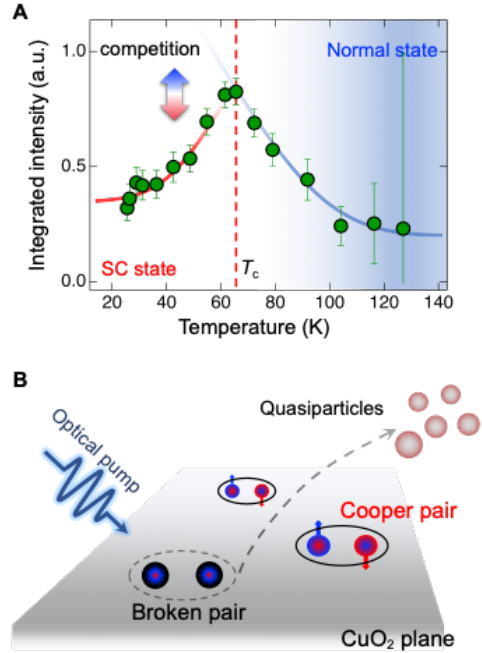

**Fig. 1. Normal and superconducting states in the CuO$_2$ plane of the YBCO.** (A) Temperature dependence of a static CDW order. It shows the integrated intensity of CDW peak at $Q = (0, q_{cdw}, \sim1.45)$ as a function of temperature, where $q_{cdw}$ is around -0.32 r.l.u. The CDW emerges below ~150 K. Below $T_c$ = 65.5 K (dashed line), the CDW peak becomes suppressed with the development of the SC. Solid lines are guides-to-the-eye. (B) A schematic sketch for the optical pump effect in a superconducting CuO$_2$ plane. The solid and dashed ellipses are a superconducting Cooper pair state and its broken state, respectively. The broken pairs transform to quasiparticles due to an optical pump.

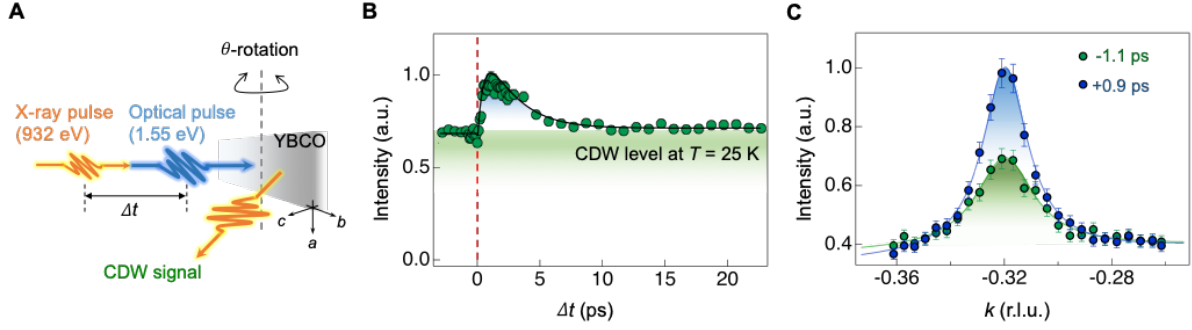

**Fig. 2. Time-resolved resonant soft x-ray scattering (tr-RSXS) on YBCO.** (A) Schematic experimental configuration for YBCO sample. The orange and blue colored pulses denote x-ray (932 eV, Cu $L_3$-edge) and optical laser (1.55 eV), respectively. θ-scans were performed to find a CDW peak at $Q = (0, q_{cdw}, 1)$. (B) CDW intensity at Q = $(0, q_{cdw}, \sim 1.45)$ as a function of Δt, measured at $T = 25$ K (below Tc). For Δt < 0, the CDW height indicates the suppressed, but non-zero, CDW due to the competition with SC. For 0 < Δt < ~7 ps, the CDW height is enhanced by the optical pump. The solid line presents a fit to exponential functions. The dashed line indicates Δt = 0. (C) The CDW peak profiles at Δt = -1.1 ps (i.e., before pumping) and Δt = +0.9 ps (i.e., maximum pumping effect on CDW). Solid lines are Lorentzian fits to the data with a linear background. Green and blue colored shades are guides-to-the-eye to denote unpumped (i.e., static) and photoinduced CDW, respectively.

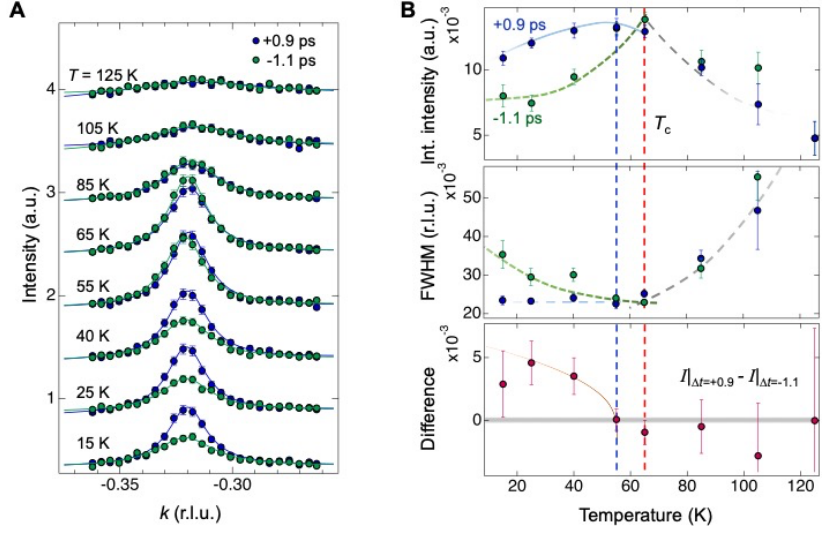

**Fig. 3. Temperature dependence of photoinduced CDW.** (A) The CDW peak profiles at (0, *k*, ~1.45) measured at Δt = +0.9 ps (blue) and -1.1 ps (green) with varying sample temperature. Solid lines are Lorentzian fits to the data with a linear background. (B) Fitted results of temperature dependence at Δt = +0.9 ps (blue) and -1.1 ps (green) for (top panel) integrated intensity and (middle panel) peak width (full-width at half maximum, FWHM). Bottom panel shows a difference of integrated intensities between Δt = +0.9 ps and -1.1 ps. Red and blue vertical dashed lines respectively represent Tc before (Δt < 0) and after (Δt = +0.9 ps) optical pumping. Solid and dashed lines are guides-to-eye. The error bars represent 1 standard deviation (s.d.) of the fit parameters.

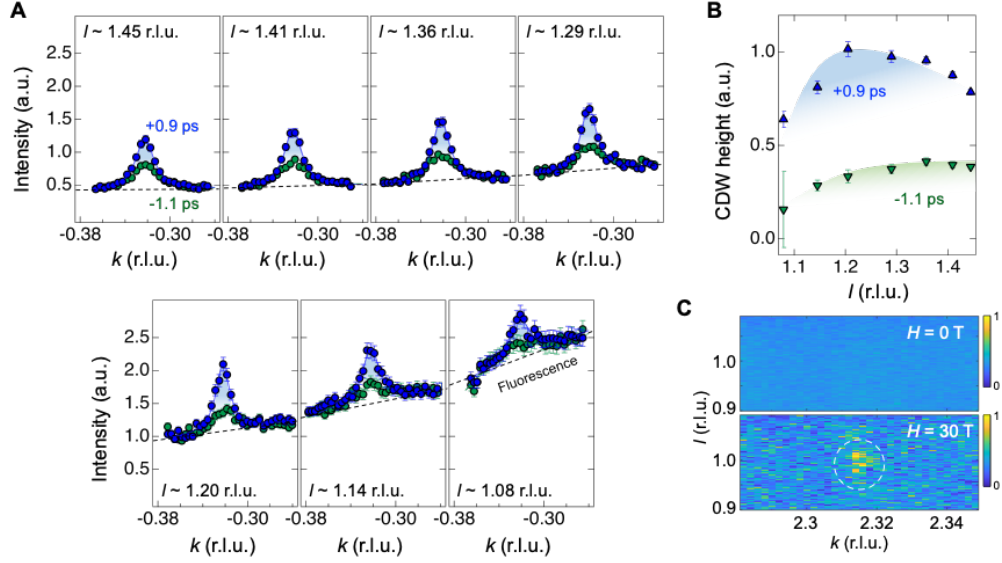

**Fig. 4. L-dependence of photoinduced CDW.** (A) The CDW peak profiles of YBCO measured at Δt = +0.9 ps and -1.1 ps with varying detector positions (i.e., *L*-dependence) and *T* = 25 K. The dashed lines indicate a fluorescence background slope at each geometrical (detector) position. As detector 2θ decreases (i.e., *l* decreases), the fluorescence background is more pronounced due to geometrical effects such as elongation of the beam footprint. Solid lines are Lorentzian fits to the data with a linear fluorescence background. (B) Fitted CDW heights with the different *l*-position (i.e., 2θ) at Δt = +0.9 ps (up-triangles) and -1.1 ps (down-triangles). Blue and green colored shades are guides-to-the-eye. The error bars represent 1 s.d. of the fit parameters. (C) The observed 3D CDW of YBCO via a pulsed magnet experiment (see Method). The field-induced CDW appears (4 shots average) around (0, 2+$q_{cdw}$, 1), compared to the zero field (80 shots average).

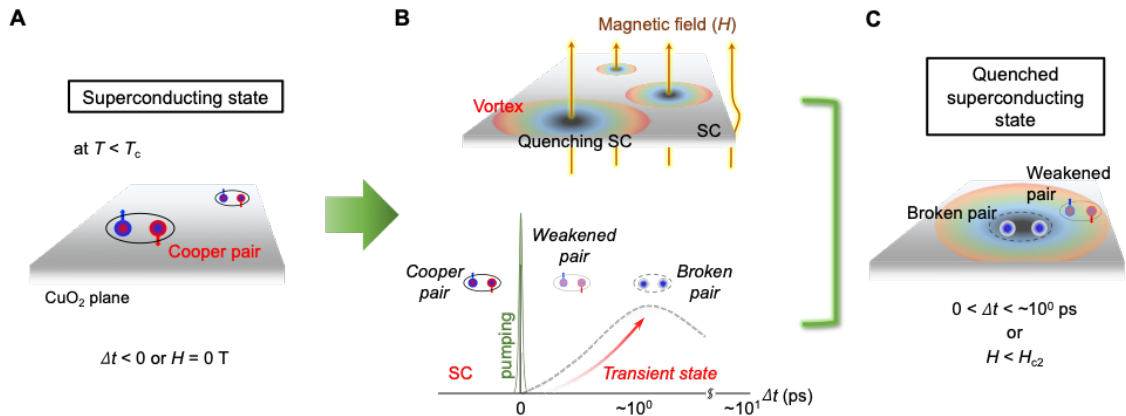

**Fig. 5. Connection between equilibrium and non-equilibrium states.** (A) Schematic cartoon of YBCO's superconducting state below $T_c$. The presence of Cooper pairs denotes SC. (B) Two independent ways to quench SC of YBCO, transforming into the normal state. Top and bottom panels are an external magnetic field approach (representing the equilibrium state) and optical pumping approach (representing the non-equilibrium state), respectively. The magnetic field approach causes the nucleation of the vortex state; the optical pumping denotes three kinds of pairing states (schematic inset: initial Cooper pair, weakened pair, and broken pair) within negative to a picosecond time frame, while SC state is transformed to a transient normal state. (C) A characterization of transient state in YBCO. The quenched SC states by either the vortex or the increased distribution of the broken (or weakened) pairs, are equivalent.

# Supplementary Materials for

## Characterization of photoinduced normal state through charge density wave in superconducting YBa$_2$Cu$_3$O$_{6.67}$

H. Jang, S. Song, T. Kihara, Y. Liu, S.-J. Lee, S.-Y. Park, M. Kim, H.-D. Kim, G. Coslovich, S. Nakata, Y. Kubota, I. Inoue, K. Tamasaku, M. Yabashi, H. Lee, C. Song, H. Nojiri, B. Keimer, C.-C. Kao, and J.-S. Lee*

*Corresponding author. Email: jslee@slac.stanford.edu

*- RSXS measurement:* The RSXS measurements were carried out at beamline 13-3 of the Stanford Synchrotron Radiation Lightsource (SSRL). The sample was mounted on an in-vacuum 4-circle diffractometer. The sample temperature was controlled by an open-circle helium cryostat. The x-ray was linearly and vertically polarized (σ-polarization), and the photon energy was tuned around Cu $L_3$-edge (932 eV). The exact $(0, k, l)$ scattering plane was aligned by the measured $(0, 0, 2)$, $(0, -1, 1)$, and $(1, 0, 1)$ structural Bragg reflections at the photon energy ~1770 eV. The scattering signal was obtained by a GaAs photodiode and normalized by the drain current on Au mesh.

*- Analysis of transient behavior of the CDW intensity:* To analyze the transient behavior of the CDW intensity on this YBCO cuprate after the optical pump, we first consider when the CDW peaks are changing. According to the CDW works in YBCO [12-14], there are three components to explain how the CDW peaks are changing: 1) Direct perturbation of Cu-charge state (e.g., photo-excited carrier); 2) Sample temperature change (heating); and 3) The known competing effect with SC (only at $T < T_c$ case). These components are respectively represented as blue-, red-, and purple-line in each panel, and the black-colored solid line shows their all summation. Note that each component consists of two exponential functions – decay and recovery with sample amplitude.

For the $T \geq T_c$ case, the temporal intensity change can be analyzed by components 1) and 2), except component 3). It is because the SC effect is not developed. In this case, as shown in Fig. S6a, the CDW firstly undergoes a fast melt near $\Delta t \sim 0$, and then the melted intensity recovers within a few ps time scales. This behavior is mainly explained by component 1). As the fluence increases, we found that the heat effect becomes pronounced with a longer time to recover to initial intensity.

Figure S6b shows an analysis of the temporal CDW intensity change under the superconducting state (i.e., $T < T_c$). It gets somewhat complicated because of the additional SC-related term (i.e., component 3). However, in this case, we could adapt the fitting parameters of components 1) and 2) analyzed at $T \geq T_c$ if the same fluence is used. For $T < T_c$, the overall temporal CDW intensity shows an enhancement because the SC becomes suppressed (Fig. 2). Similar to the case at $T = T_c$ (see Fig. S6a), the higher fluence induces more heat effect and slower the CDW intensity returning. Interestingly, the intensity maximized $\Delta t$ becomes delayed as the fluence increases, i.e., from ~1 ps with 15 μJ/cm$^2$ to ~3 ps with 60 μJ/cm$^2$. We analyze this also comes from the pronounced heat effect in the higher fluence range. In order to minimize the heat effect during this study, we employed a mild pump fluence ~15 µJ/cm$^2$ in the tr-RSXS measurements.

**- X-ray and optical pump penetration depth:** According to CXRO database (The Center for X-Ray Optics at Lawrence Berkeley National Laboratory, https://www.cxro.lbl.gov/), the normal incidence penetration depth around Cu $L_3$-edge is estimated to be about 120 nm. Considering the sample angle ($\theta$ = 46° for $L$ = 1.46) and the detector angle ($2\theta$ = 167°), the RSXS probing depth is estimated as ~46 nm from the sample surface. According to the previous reports Refs. [35 (L. Stojchevska *et al.*) and 44 (B. Liu *et al.*)], the penetration depth of 800-nm pump laser in YBa$_2$Cu$_3$O$_{6+x}$ is estimated as the range of 80-200 nm.

**- X-ray polarization effect in L-dependence of tr-RSXS measurement:** As shown in Fig. 2a, the incident x-ray polarization signal is the *pi*($\pi$)-polarization. Since the CDW intensity is proportional to $\pi\ ' \cdot \pi = \cos(2\theta)$, the intensity was rescaled by multiplying $1/\cos(2\theta)$ to compare the intensities in *L*-dependence.

**- Optical laser polarization effect in tr-RSXS measurement:** In this study, we applied the *p*-polarized optical pump laser (i.e., *p* // *bc*-plane). Since the *p*-polarized laser has both components along the parallel and orthogonal direction to the CuO$_2$ plane, a pumping effect in the CuO$_2$ plane via the *p*-polarization is effectively weaker than that via the *s*-polarization (*s* // *ab*-plane). As shown in Fig. S7, the optical laser polarization effects were explored. We confirmed that the *p*-polarization needs more fluence to generate the similar effect shown the *s*-polarization, especially while the fluence value is high. For the fluence = 15 µJ/cm$^2$, which was mainly used in this study, we could find that the difference between two polarizations is minimized.

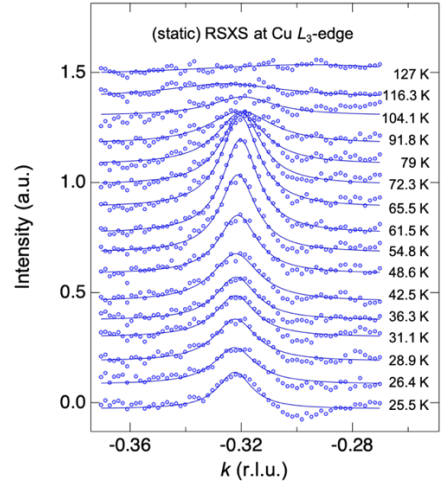

**fig. S1. Temperature dependence of the static CDW order in YBCO.** The CDW peak observed at ***Q*** = (0, $q_{cdw}$, ~1.45) where $q_{cdw}$ is around -0.32 reciprocal lattice unit (r.l.u.). The data are subtracted by high temperature (~150 K) data. Solid lines are Lorentzian fits to the data.

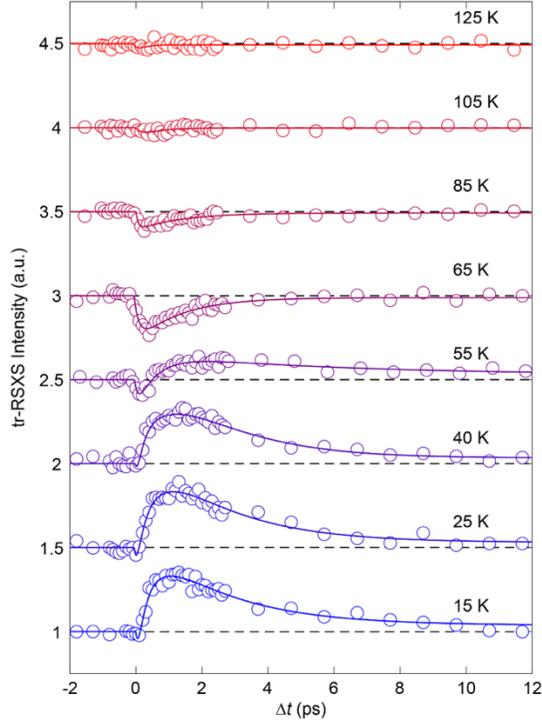

**fig. S2. Temporal behavior of the CDW order as a function of temperature**. The tr-RSXS intensity was recorded at the CDW at $Q$ = (0, -0.32, ~1.45). Each delay data has been vertically shifted for clarity. The solid line presents fitting by exponential functions. The dashed lines are baselines determined from the negative delay region.

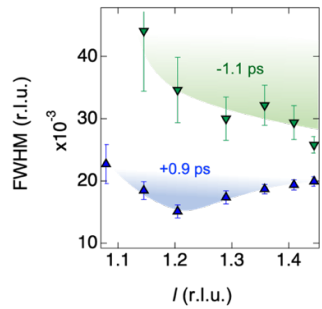

**fig. S3. *L*-dependence of photoinduced CDW**. Fitted FWHM of the CDW peaks at *l* positions (i.e., $2\theta$) of $\Delta t$ = +0.9 ps (up-triangles) and -1.1 ps (down-triangles). Blue and green colored shades are guides-to-the-eye. The error bars represent 1 standard deviation (s.d.) of the fit parameters.

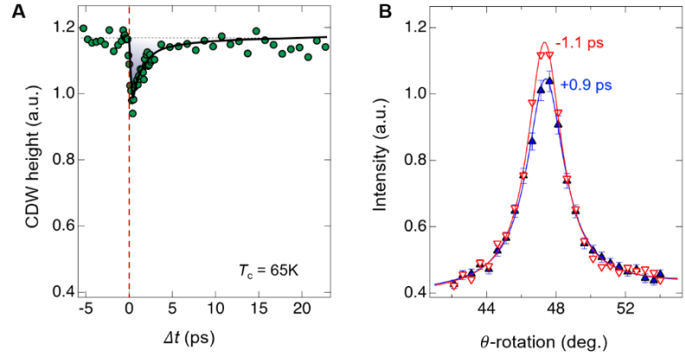

**fig. S4. Photoinduced CDW behavior.** (**A**) CDW intensity at $Q = (0, -0.32, 1.45)$ as a function of $\Delta t$, measured at $T = 65$ K ($\sim T_c$). For $\Delta t < 0$, the data indicate the CDW intensity before the pump. For $0 < \Delta t < \sim 5$ ps, the CDW is melted by the optical pump and then is recovered. The maximum melting of CDW occurs at $\Delta t \sim 0.2$ ps. The dashed lines are guides-to-the-eye. (**B**) The CDW peak profiles at $\Delta t = -1.1$ ps and $\Delta t = +0.9$ ps. At $\Delta t = 0.9$ ps, the CDW intensity recovered to about 83 % of the intensity at $\Delta t < 0$. Solid lines are Lorentzian fits to the data with a linear background.

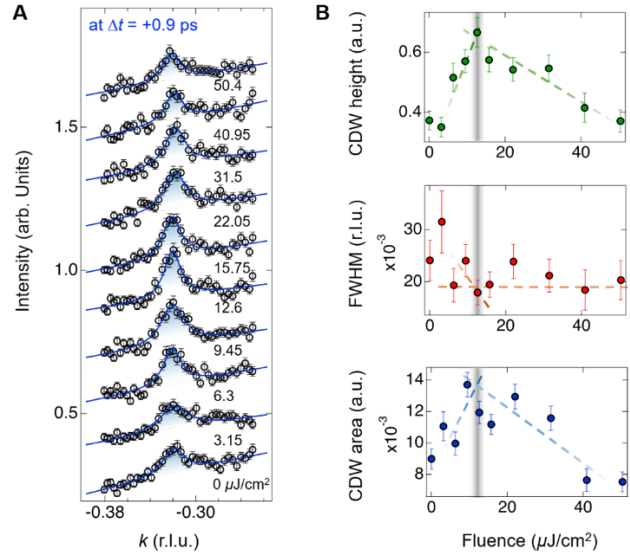

**fig. S5. Fluence dependence of photoinduced CDW.** (**A**) CDW intensity at $Q = (0, -0.32, 1.45)$ with varying laser fluence, measured at $T = 25$ K and $\Delta t = +0.9$ ps. The blue colored lines are Lorentzian fits to the data with a linear background. Each fluence data are vertically shifted by $\sim$ 0.3 arb. Units. (**B**) Fitted CDW peaks with the different fluence – top: CDW peak's height, middle: FWHM, bottom: integrated intensity (CDW area). The error bars represent 1 standard deviation (s.d.) of the fit parameters. Gray colored shade denotes the employed fluence value in the tr-RSXS. All dashed lines are guides-to-the-eye.

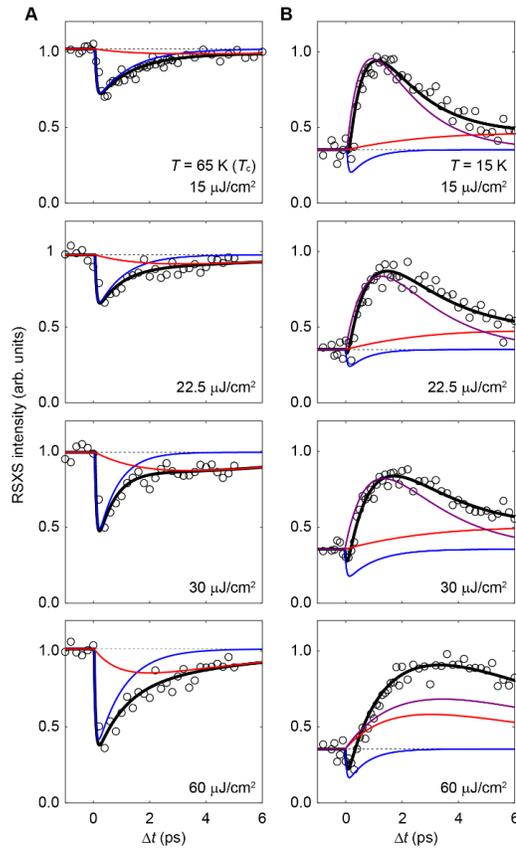

**fig. S6. Fluence-dependent delay scans of CDW**. Data taken at (0, -0.32, 1.45) and at (**A**) 65 K ($T_c$), (**B**) 15 K. From top to bottom panels, incident fluence increases. Blue, red, and purple lines present fitting of direct perturbation, heating, and SC effect, respectively. Black solid lines show their summation. Black dashed lines display the intensity at $\Delta t < 0$.

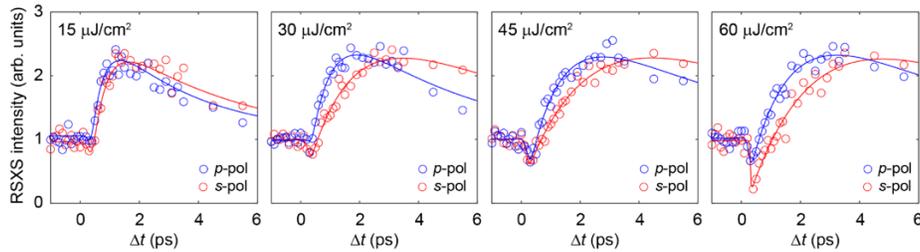

**fig. S7. Optical pump laser polarization effect of delay scans of CDW**. Data taken at (0, -0.32, 1.45) and 20 K. *p*-pol (*s*-pol) is parallel to the crystalline *bc* (*ab*)-plane. The solid lines present exponential fitting.

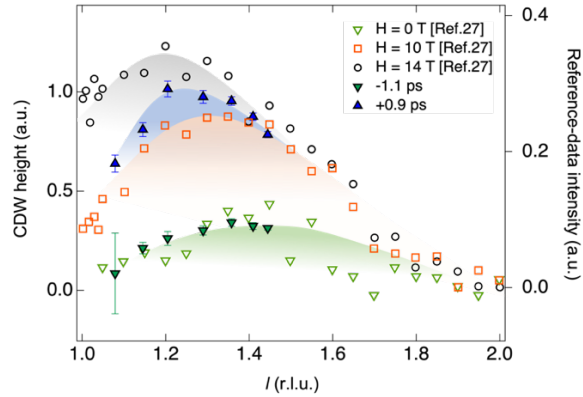

**fig. S8. Comparison of *L*-dependence between the pump-probe and the magnetic field.** The magnetic data were taken from the Ref. [27] (Chang *et al.*). To superpose two different data set, we converted *x*-axis of the reference ($x_{ref}$) as follow: $l = 2 - x_{ref}$. Also, it is important to note that the magnetic (*H*) field data in the reference were measured at $T = 22$ K (this study: $T = 25$ K). Also, the *H*-field study was measured by the non-resonant hard x-ray scattering (this study by resonant soft x-ray scattering).

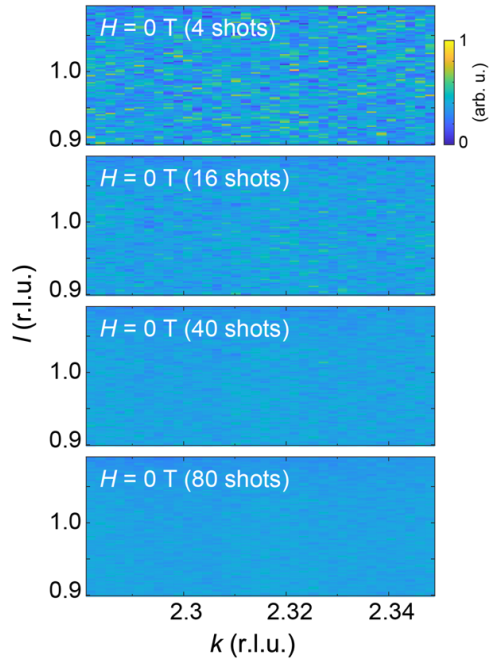

**fig. S9. Noise level comparison depending on the number of shots in the magnetic field experiment.** All panels show *kl*-space around $(0, 2+q_{cdw}, 1)$ under $H = 0$ T. The panels (from top to bottom) are 4-, 16-, 40-, and 80-shots average, respectively. With increasing the shot acquisition, the noise level is reduced.